## MATERIALS SCIENCE

# Three-dimensional self-assembly using dipolar interaction

Leon Abelmann[1,2]*, Tijmen A. G. Hageman[1,2], Per A. Löthman[1,2], Massimo Mastrangeli[3], Miko C. Elwenspoek[2]

Interaction between dipolar forces, such as permanent magnets, generally leads to the formation of one-dimensional chains and rings. We investigated whether it was possible to let dipoles self-assemble into three-dimensional structures by encapsulating them in a shell with a specific shape. We found that the condition for self-assembly of a three-dimensional crystal is satisfied when the energies of dipoles in the parallel and antiparallel states are equal. Our experiments show that the most regular structures are formed using cylinders and cuboids and not by spheroids. This simple design rule will help the self-assembly community to realize three-dimensional crystals from objects in the micrometer range, which opens up the way toward previously unknown metamaterials.

## INTRODUCTION

Crystal growth is a form of self-assembly (*1*–*3*), where the individual objects (atoms and molecules) arrange into regular arrays. The process of crystal formation has been studied in great detail (*4*) on a vast range of materials and has a widespread technological impact ranging from silicon single crystals (*5*) for the semiconductor industry to diffraction studies on proteins (*6*). Crystal growth takes place by a nucleation and growth mechanism. Nucleation starts on well-defined templates (epitaxy) (*7*) or random imperfections (formation of snowflakes) or occurs spontaneously in space (*8*). The latter is the subject of this study.

Crystal formation of objects larger than atoms and molecules is receiving increasing attention (*9*, *10*), driven by the promise of metamaterials with novel functionality (*11*, *12*). There are beautiful examples of crystal growth from silica or polymer spheres, such as for three-dimensional (3D) photonic crystals (*13*–*15*). In these examples, the self-assembly process relies on the evaporation of a solvent to bring the components in each other's vicinity, possibly assisted by solvent flow (*16*). In simulations, the increase in particle concentration is often modeled by slowly contracting the simulation space (*17*). Alternatively, self-assembly can be driven by sedimentation (*18*). These approaches generally lead to close-packed structures (*10*). After solvent evaporation, the assembly is held together by van der Waals forces between particles or by residues from the solvent (cementing) (*19*). van der Waals forces act over a short range and become less effective for larger objects. Therefore, long-range static forces are being investigated, such as in a binary mixture of oppositely charged spheres (*20*).

When growing crystals of identical objects, the objects themselves obviously cannot have a net charge. On the microscale, one can use induced or permanent dipoles, which could either be of electrostatic or magnetic origin (*21*). The dipole moments can be either induced by an externally applied field (*22*, *23*) or permanent. In this study, we investigate the possibility of self-assembling crystals using permanent magnetic dipolar forces. Permanent magnetic dipoles are especially useful for objects of large size, since magnetic poles are not easily screened. What we learn in this way from magnetic dipoles can be applied to electric dipoles, since long-range forces between both types of dipoles are identical.

Magnetic dipoles allow us to increase the object size to the millimeter range (*24*). At this scale, it is easy to study the process of 3D self-assembly in real time. By doing so, we obtain information not only on the final product of self-assembly but also on the processes that lead to the formation of the assembly. Real-time observation of the self-assembly process provides clues on the origin of defects.

We performed experiments with millimeter-sized permanent magnets, embedded in a polymer shell of varying shape. The objects were submerged in water, in a transparent conical cylinder with an inner diameter ranging from 9 to 19 cm (Fig. 1A). Gravitational forces were counterbalanced by an upward water flow that decreased in speed because of the conical shape of the boundary, so that the objects remain in the field of view of the camera. The adjustable turbulence in the flow created disturbing forces to enable the system to reach the global energy minimum. These disturbing forces provide stochastic kinetic energy to the objects, leading to a motion analogous to Brownian motion (*25*).

The interaction between permanent spherical dipoles results in the formation of chains (*26*). Figure 1B shows an example with eight dipoles that line up in a ring. The formation of these rings is well understood (27–29). The dipolar forces first organize the spheres into a line. The energy of this configuration, relative to the energy of a dipole pair ($N = 2$), is

$$u_N^{\text{line}} = \frac{-2}{N} \sum_{i=1}^{N-1} \frac{N-i}{i^3}$$

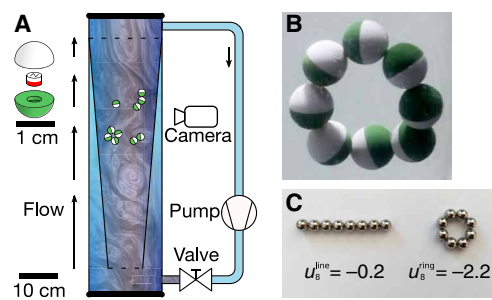

**Fig. 1. The self-assembly experiment.** (**A**) 3D printed polymer objects with embedded permanent magnets were inserted in a transparent cylinder with an upward flow. The flow counteracts the drop velocity of the objects, and the flow's turbulence provides a disturbing force. A tapered transparent insert was used to provide a gradient in the flow velocity, which ensured that the objects levitate in front of the video cameras. (**B**) Spherical objects form linear chains. When eight spheres are inserted in the flow, the most stable configuration is a circle, which has 10% lower energy than a linear chain (**C**). Photo credit: L. Abelmann (Saarland University and University of Twente).

[1]KIST Europe, Saarland University, Saarbrücken, Germany. [2]University of Twente, Enschede, Netherlands. [3]Electronic Components, Technology and Materials, Department of Microelectronics, Delft University of Technology, Delft, Netherlands.
*Corresponding author. Email: l.abelmann@utwente.nl

For more than three spheres, a lower energy state can be reached by closing the line into a ring

$$u_N^{\text{ring}} = \frac{-1}{4}\sin^3\left(\frac{\pi}{N}\right) \sum_{k=1}^{N-1} \frac{3 + \cos(2\pi k/N)}{\sin^3(\pi k/N)}$$

In case of eight spheres, the energy gain is substantial (Fig. 1C), so the ring forms easily and remains intact.

These 1D chains form because the antiparallel dipole configuration has twice the energy of the parallel configuration at identical dipole center-to-center distance (Fig. 2A, left). To achieve assemblies with higher dimensionality, we can use the shape of the polymer shell to change the distance between the dipoles for different orientations. By elongating the shell, we can increase the distance between the dipole centers in the parallel configuration to the point that the energy of the antiparallel configuration is lower than the parallel configuration. In this case, the antiparallel configuration is preferred, and we obtain 2D plate-like structures (Fig. 2A, center). If the energies of the parallel and antiparallel states are equal, then newly arriving dipoles align both in a parallel and an antiparallel fashion, and one would expect 3D structures (Fig. 2A, right).

### RESULTS

We demonstrated this strategy for eight spheroids, cylinders, and cuboids. The energy difference between the antiparallel and parallel states was chosen to be 40 μJ for all shapes (Fig. 2B, first column). As predicted, we observe the formation of line structures. Only a spherical shell allows the formation of a ring. Cylinders and cubes form rigid lines. For the cubes, this is in agreement with molecular dynamics studies and experiments on iron nanocubes prepared in a gas-phase cluster gun (30). By reversing the energy difference between the parallel and antiparallel states, so that the antiparallel state has the lowest energy (Fig. 2B, center column), we observed clear plate structures for the cylinders, less perfect plates for the cuboids, and irregular structures for the spheroids. When both energies were equal (Fig. 2B, third column), the cylinders started to form perfect 3D 2 by 2 by 2 clusters (red circle in Fig. 2B) The cuboids' assemblies suffered from relatively stable attachments of cuboids at a 90° orientation, which led to magnetic flux closure and prohibited further growth. The spheroids formed a complex double-ring structure, which resembled the prediction made by Messina et al.0(29) for larger numbers of objects.

In our experiment, the structures of spheroids stay together for several minutes. This is much longer than is the case for the structures of cylinders and cubes, which often break up into parts after a few seconds. The ring structure of spheres breaks up rather easily into a chain but then reconnects again into a ring in less than a minute. We believe that the higher stability of the spheroid structures is caused by their ability to misalign without immediately increasing their distance, which decreases the force between the magnets. In general, the chain structures break up more easily (within a few seconds) than the plates or crystals. This is expected since breaking a chain only requires to break a single bond between two objects, whereas for plates and crystals multiple bonds need to be broken simultaneously. In addition, the cylinders and cubes form rigid chains that are very long, resulting in frequent contact with the reactor walls and breaking of the chain.

Out of the shapes that we investigated, cylinders appear to be most suited for self-assembly into well-defined 3D structures. Experiments with an increasing number of objects (fig. S1) confirmed that spheroids do not form regular crystals, in contrast to cylinders and cuboids. Insights as to why this happens can be obtained by studying the process of self-assembly itself (movies S1 and S2). The spheroids tend to stay together longer as a cluster than the cylinders and cuboids. Clusters of cylinders and cuboids often break up into two smaller clusters, which then realign to form a more regular crystal.

The breakup of assemblies happens more often for larger assemblies, probably because shear forces tend to increase with assembly size. This effect might be amplified by our former observation that the energy in the turbulent flow increases with increasing length scale (25). We are not sure whether this is a general aspect of turbulent-driven self-assembly or a particular aspect of our experimental configuration. This question needs further investigation, for instance by changing the absolute size of the objects.

In particular, single cylinders attached to the cluster can wander rather easily over the surface, which is not the case for cuboids. The cuboids take a longer time to attach to each other. We suspect that, to fully adhere, the water between the cuboids needs to be pushed out over a few millimeters. In the case of spheroids and cylinders, the amount of water to be displaced is far less.

### DISCUSSION

These experiments demonstrate that 3D structures can self-assemble from dipolar forces, provided that there is no preference for parallel or antiparallel alignment. This can be achieved by balancing dipolar forces with steric interactions induced by the specific shape of the object. It is interesting that the shape of the object plays such a major role. The spheroids have many orientations under which they can attach to a forming cluster; the cuboids, on the other hand, only have a few. The cylindrical shape appears to be a good compromise. In addition, in 2D self-assembly, it was shown that a rounded shape helps to achieve regular crystals (*31*).

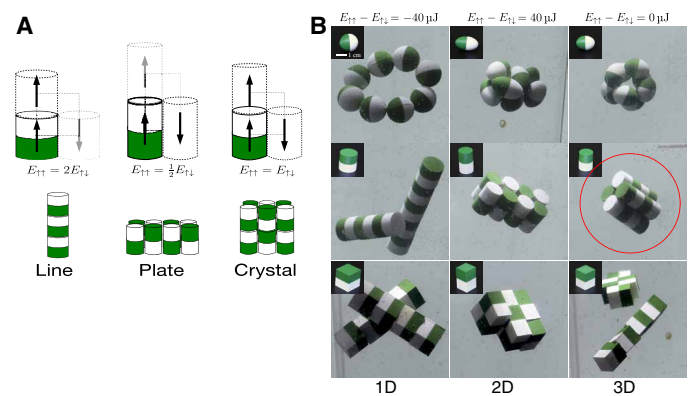

**Fig. 2. 3D self-assembly of dipoles.** (**A**) Equally spaced dipoles prefer parallel alignment (black arrows). By elongating the shape of the shell around the dipoles, we can favor the antiparallel configuration, so that plates of objects assemble. When the energy of the parallel and antiparallel configuration is exactly equal, we expect 3D crystals. (**B**) This strategy works best with cylindrical objects. From left to right, we varied the shape so that the energy of the parallel configuration is twice (left), half (center), and exactly equal (right) to that of the antiparallel configuration. The red encircled assembly of cylinders (middle row) is a regular 3D 2 by 2 by 2 cluster. The cylindrical objects in the second row reproduced the plate prediction of (A). The spheroids (top row) and the cubes (bottom row) exhibited line structures in the first column but more complex behavior when their shape was adjusted.

As well as the energy difference between the final states, the paths toward those energy minima are also of major importance. This observation is in agreement with molecular dynamic simulations, which show that spheres are more likely to form larger clusters than cubes (*32*) and that dipolar interaction disturbs the crystal formation of cubes (*33*).

These results encourage experiments on self-assembly of crystals at the microscale using permanent magnetic dipoles. The millimeter-sized cylindrical objects could be miniaturized by lithographic techniques and anisotropic etching on magnetic thin films with a perpendicular easy axis sandwiched between two nonmagnetic films, such as is currently used in magnetic random access memories (*34*). From there, one can envision interesting metamaterials, such as artificial antiferromagnets, piezo-magnetic materials with a negative Poisson ratio (*11*), or 3D magnetic ring-core memories (*35*).

The forces between dipoles do not change when we reduce the size of the dipoles, apart from a scaling factor. Neither does it matter whether the dipoles are of magnetic or electrical origin. This implies that we can generalize the outcome of these experiments to the design of electrostatically interacting objects of micrometer size for 3D self-assembly, aimed at applications such as photonic crystals (*14*), supermaterials (*11*), 3D electronics (*36*), or memories (*35*).

## MATERIALS AND METHODS
The experimental setup was introduced and characterized in (*24*, *25*). New to this setup was a cone-shaped inset, which created a flow gradient meant to center particles in the middle and prevent interaction with the top and bottom. The 3D printed shells are spheroids, cylinders, or cuboid, ordered in increasing extent to which the particle poses geometrical restrictions on how they can connect. All objects have an identical cylindrical, 4 mm by 4 mm axially magnetized NdFeB core and are color-coded on the basis of polarization. They are designed such that $E_{ax} - E_{diam} \in \{-40, 0, 40\}$ µJ, in order of increasing aspect ratio, while ensuring that their minimum connection energy stays at −80 µJ. The 3D design files (STL format) for the shells are available in the Supplementary Materials. The dimensions of the objects, measured with a caliper, are listed in table S1. Objects in various amounts (*8*, *12*, *16*) were inserted into the reactor with appropriate flow speed (approximately 9 cm/s) settings to create neutral buoyancy.

## SUPPLEMENTARY MATERIALS
Supplementary material for this article is available at http://advances.sciencemag.org/cgi/content/full/6/19/eaba2007/DC1

**Acknowledgments:** We would like to acknowledge M. Marsman, L. Woldering, and R. Sanders for conception and realization of the setup; A. Manz and G. Krijnen for valuable advice and discussions, Proof-Reading-Service.com for correcting the manuscript; and the anonymous reviewers for improving the manuscript and suggesting a more elegant summation for the first equation. **Funding:** This work was funded by KIST Europe, under basic grant 11908. **Author contributions:** L.A., T.A.G.H., and M.C.E. generated the concept to use shape to tailor the dipolar interaction. T.A.G.H. and P.A.L. performed the experiments. All authors contributed substantially to the manuscript. **Competing interests:** The authors declare that they have no competing interests. **Data and materials availability:** All data needed to evaluate the conclusions in the paper are present in the paper and/or the Supplementary Materials. Additional data related to this paper may be requested from the authors.

Submitted 13 November 2019
Accepted 24 February 2020
Published 8 May 2020
10.1126/sciadv.aba2007

**Citation:** L. Abelmann, T. A. G. Hageman, P. A. Löthman, M. Mastrangeli, M. C. Elwenspoek, Three-dimensional self-assembly using dipolar interaction. *Sci. Adv.* **6**, eaba2007 (2020).

# Science Advances

## Three-dimensional self-assembly using dipolar interaction

Leon Abelmann, Tijmen A. G. Hageman, Per A. Löthman, Massimo Mastrangeli and Miko C. Elwenspoek